\documentstyle[aps,prb]{revtex}

\begin{document}
\draft

\title{Hysteretic current-voltage characteristics and resistance switching 
at an epitaxial oxide Schottky junction 
SrRuO$_{3}$/SrTi$_{0.99}$Nb$_{0.01}$O$_{3}$}

\author{T. Fujii and M. Kawasaki}
\address{Correlated Electron Research Center (CERC), 
National Institute of Advanced Industrial Science and Technology (AIST), 
Tsukuba, Ibaraki 305-8562, Japan\\
and Institute for Materials Research, Tohoku University, Sendai 980-8577, Japan}

\author{A. Sawa\cite{a} and H. Akoh}
\address{Correlated Electron Research Center (CERC), 
National Institute of Advanced Industrial Science and Technology (AIST), 
Tsukuba, Ibaraki 305-8562, Japan\\
and CREST-JST, Kawaguchi, Saitama 322-0012, Japan}

\author{Y. Kawazoe}
\address{Institute for Materials Research, Tohoku University, 
Sendai 980-8577, Japan}

\author{Y. Tokura}
\address{Correlated Electron Research Center (CERC), 
National Institute of Advanced Industrial Science and Technology (AIST), 
Tsukuba, Ibaraki 305-8562, Japan\\
and Department of Applied Physics, University of Tokyo 113-8658, Japan}

\date{\today}
\maketitle

\begin{abstract}
Transport properties have been studied for a perovskite heterojunction 
consisting of SrRuO$_{3}$ (SRO) film epitaxially grown on 
SrTi$_{0.99}$Nb$_{0.01}$O$_{3}$ (Nb:STO) substrate. 
The SRO/Nb:STO interface exhibits rectifying current-voltage 
($I$-$V$) characteristics agreeing with those of a Schottky junction composed 
of a deep work-function metal (SRO) and an $n$-type semiconductor (Nb:STO). 
A hysteresis appears in the $I$-$V$ characteristics, where high resistance and 
low resistance states are induced by reverse and forward bias stresses, 
respectively. 
The resistance switching is also triggered by applying short 
voltage pulses of 1 $\mu$s - 10 ms duration. 

\end{abstract}


Recently, reversible resistance switching between two or multilevel 
resistance states has been found to take place by short voltage pulses 
at room temperature in capacitor-like devices composed of a wide variety 
of insulating perovskite oxides such as 
manganites,\cite{ref-1,ref-2,ref-3} titanates,\cite{ref-4}
and zirconates\cite{ref-5} sandwiched between two metallic electrodes. 
This resistance switching attracts considerable attention due 
to the potential for device application such as resistance random 
access memories (RRAM).\cite{ref-6}
The origin of resistance switching, however, is still an open question. 
One of the possibilities is the bulk effect\cite{ref-1,ref-4,ref-5,ref-6} 
that a phase transition of perovskite takes place between insulating 
and conducting states, 
similar to the breakdown of charge-ordered insulating state in manganites 
induced by electric-field at low temperature.\cite{ref-7,ref-8,ref-9} 
The other is the interface effect, where voltage pulses reversibly 
alter the nature of potential barrier formed in the insulating 
(or semiconducting) perovskite in contact with metallic 
electrodes.\cite{ref-2,ref-3}
We have recently shown that the resistance switching occurs at a 
Ti/Pr$_{0.7}$Ca$_{0.3}$MnO$_{3}$ (PCMO) interface,\cite{ref-3} 
which exhibits Schottky-like current-voltage ($I$-$V$) characteristics, 
where Ti and PCMO can be regarded as a shallow work-function metal 
and a $p$-type semiconductor, respectively. 
A possible origin for the resistance switching is attributed 
to the change in Schottky barrier height (or width) by trapped 
charge carriers at the interface states. However, 
non-epitaxial structure and chemically incompatible materials combination 
make it difficult to characterize the transport properties and 
interface electronic structure in detail.

In the present study, we have investigated the transport properties of 
a heteroepitaxial perovskite oxide interface consisting of SrRuO$_{3}$ (SRO) 
deposited on (001) SrTi$_{0.99}$Nb$_{0.01}$O$_{3}$ (Nb:STO) single 
crystal substrate. 
The SRO/Nb:STO interface exhibits rectifying Schottky-like $I$-$V$ 
characteristics with large hysteresis and the resistance can be changed 
by applying pulsed-voltage stress.

Epitaxial SRO thin films (100 nm) were grown on (001) Nb:STO single 
crystal substrates by a pulse laser deposition technique. 
Typical growth conditions were a substrate temperature of 700 $^{\circ}$C 
and an oxygen pressure of 100 mTorr. 
After the deposition, the films were $in$-$site$ annealed at 400 
$^{\circ}$C for 30 minutes under an oxygen pressure of 500 Torr 
and then cooled down to room temperature. 
The crystal structure was analyzed by a four-circle x-ray diffractometer. 
The full-width at half maximum of the (002) rocking curve for SRO films is 
as narrow as 0.04 $^{\circ}$. 
The in-plane lattice is compressively strained to match coherently with that 
of the substrate, resulting in a tetragonal lattice structure with an 
out-of-plane lattice constant of 0.395 nm.\cite{ref-10}

The lower inset in Fig. \ref{fig-1} illustrates the sample structure and the 
measurement configuration. 
A gold electrode layer (230 nm) was deposited $ex$-$situ$ on SRO by 
electron-beam evaporation. 
Mesa-structures with 100 $\mu$m $\times$ 100 $\mu$m in size were fabricated by 
conventional photolithography and Ar ion etching. 
The $I$-$V$ characteristics were measured by the three-probe method using 
very large mesa (several mm$^2$) as a quasi-ohmic voltage contact to Nb:STO. 
Positive bias is defined as the current flow from the Nb:STO to SRO. 
Typical $I$-$V$ characteristics are shown in Fig. \ref{fig-1}. 
The SRO/Nb:STO junction exhibits rectifying characteristics, 
which agree with the ones expected for a Schottky junction employing 
SRO and Nb:STO as a deep work-function metal and an $n$-type semiconductor, 
respectively. 
In the positive bias region, i.e., the reverse bias of the Schottky junction, 
the current scarcely flows at low bias and increases with increasing 
bias voltage above 2 V. 
Sweeping back to the negative voltage, i.e., the forward bias of 
the Schottky junction, the current suddenly increases at a certain 
voltage which corresponds to the flat-band condition or turn-on voltage. 
By reducing the temperature to $T$ = 50 K, the turn-on voltage slightly 
increases, which agree with the temperature dependence of a typical 
Schottky junction.

The $I$-$V$ characteristic at 300 K is plotted in a semilogarithmic 
current scale as solid lines in Fig. \ref{fig-2}(a). 
There are large hystereses in $I$-$V$ characteristics at low bias 
voltage both in forward and reverse bias regions, which are not apparent in 
liner scale ones (Fig. \ref{fig-1}). 
Here, we define high resistance state (HRS) and low resistance state 
(LRS) for lower and upper branches of the hysteretic $I$-$V$ characteristic, 
respectively. 
As can be seen, the forward bias scan to -1.4 V drives the junction to 
LRS and the reverse bias scan to 4 V turned the junction to HRS. 
LRS and HRS were kept unchanged when voltage polarity is changed through 
0 V ($(1) \to (2) \to (3) \to (4)$). 
The $I$-$V$ curve at HRS in the forward bias region (4) is nearly 
a straight line, agreeing with a Schottky barrier model; forward 
bias current passing through a barrier with a height of $\phi_{\rm B}$ 
is proportional to 
$\exp{[-e(\phi_{\rm B} - \mid V  \mid)/k_{\rm B}T}]$, 
where $e$ is electron charge and $k_{\rm B}$ BoltzmannÕs 
constant (see upper inset in Fig. \ref{fig-1}). 
Although the junction gives rather leaky $I$-$V$ characteristics in 
the reverse bias region, we discuss the hysteretic behavior and 
resistance switching hereafter with assuming that the SRO/Nb:STO junction 
is a Schottky junction. 
Therefore, the LRS can be regarded as possessing a reduced barrier 
height and/or an opened extra parallel current path such as tunneling.

Now, we examine the evolution of hysteresis by widening the span of 
the voltage scan. 
In Fig. \ref{fig-2}(a), the bias voltage was swept as 
-1.4 V $\to$ 0 V $\to$ $V_{\rm max}$ $\to$ 0 V $\to$ -1.4 V 
with $V_{\rm max}$ varied as 1, 2, 3, and 4 V, where the junction was reset 
to LRS at the initial stage (see branches of (1) and (2)). 
When $V_{\rm max}$ was 1 V, the LRS was hardly converted to HRS. 
As the $V_{\rm max}$ was increased from 2 to 4 V, the junction was transformed 
gradually to HRS (see branches (3) and (4)) and the hysteresis was opened, 
resulting in the current differences over two orders of magnitude in 
low voltage regions. 
In Fig. \ref{fig-2}(b), the bias voltage was swept as 
4 V $\to$ 0 V $\to$ $V_{\rm min}$ $\to$ 0 V $\to$ 4 V 
with $V_{\rm min}$ varied as -0.8, -1.0, -1.2, and -1.4 V. 
In a similar manner to the cases of Fig. \ref{fig-2}(a), initial HRS was 
gradually converted to LRS by increasing the voltage span in forward bias. 
Therefore, the barrier height reduction and/or an opening of additional 
current path can be developed by higher forward bias stress and they 
can be diminished by higher reverse bias stress. 

When the junction was stressed by voltage pulses, the resistance was 
switched between rather steady HRS and variable LRS. 
The experiments were carried out as follows. 
First, the junction was brought into HRS by scanning the voltage to 
high enough reverse bias ($V_{\rm max}$ = 4 V at 300 K and 50 K) as shown in 
Fig. \ref{fig-2}(b). 
Then, voltage pulses of $V_{\rm p}$ = $\pm$10 V were applied with a duration 
of $\tau_{\rm p}$. 
The resistance of the junction was evaluated by reading the current at 
a voltage bias ($V_{\rm bias}$) of -0.3 V (300 K) or -0.7 V (50 K) as defined 
in Fig. \ref{fig-3}(a). 
Extreme resistance values corresponding to the resistances with 
$\tau_{\rm p} \to \infty$ were measured during $I$-$V$ scans after the 
junction was biased to $V$ = 4 V for HRS and to $V$ = -1.4 V (300 K) 
or -1.6 V (50 K) for LRS. 
These values were defined as $R_{\rm H}^0$ and $R_{\rm L}^0$ as shown 
in Fig. \ref{fig-3}(a) and plotted on the right ordinates in Figs. \ref{fig-3}(b) and (c). 
When $\tau_{\rm p}$ = 10 ms at 50 K, the resistance switching between HRS 
and LRS took place between almost $R_{\rm H}^0$ and $R_{\rm L}^0$ 
at the first pulse and additional application of pulses scarcely 
induced further change of resistance. 
As $\tau_{\rm p}$ was decreased, the resistance at LRS ($R_{\rm L}$) increased, 
while the resistance at HRS ($R_{\rm H}$) stayed constant close to $R_{\rm H}^0$. 
The ratio $R_{\rm H}$/$R_{\rm L}$ was reduced from $10^2$ at 
$\tau_{\rm p}$ = 10 ms to 2 at $\tau_{\rm p}$ = 1 $\mu$s. 
At 300 K, the $\tau_{\rm p}$ dependence of resistance switching showed a similar 
tendency (Fig. \ref{fig-3}(c)) to that at 50 K (Fig. \ref{fig-3}(b)) 
but LRS became more unstable. 
At $\tau_{\rm p}$ = 1 $\mu$s, $R_{\rm L}$ was close to $R_{\rm H}^0$ 
and switching was hardly seen. 
Even at $\tau_{\rm p}$ = 10 ms, more than ten pulses were needed to establish 
steady $R_{\rm L}$ value and $R_{\rm L}$ is still much higher than 
$R_{\rm L}^0$. 
Nonetheless, the switching direction, LRS $\to$ HRS upon reverse 
$V_{\rm p}$ and $vice$ $versa$, are the same to those appeared in $I$-$V$ scan 
experiments (Fig. \ref{fig-2}).

So far, similar rectifying $I$-$V$ characteristics have been seen 
in heteroepitaxial perovskite junctions composed of Sr(Ti,Nb)O$_{3}$ and 
various metallic perovskites such as 
YBa$_{2}$Cu$_{3}$O$_{7}$,\cite{ref-11} 
(Ba,K)BiO$_{3}$,\cite{ref-12} and (La,Sr)MnO$_{3}$.\cite{ref-13}
In all cases, the rectification direction is consistent with that 
of a Schottky junction with regarding Sr(Ti,Nb)O$_{3}$ and 
metallic perovskites as $n$-type semiconductor and deep work-function metals, 
respectively. 
The present SrRuO$_{3}$/Sr(Ti,Nb)O$_{3}$ junction is considered to 
be one of them. 
However, such distinct hysteretic behaviors as seen in Fig. \ref{fig-2}
have never been reported. 
Previously, we found that Ti/PCMO junctions showed rectifying 
$I$-$V$ characteristics with just opposite combination of materials, 
namely shallow work-function metal and $p$-type 
semiconductor.\cite{ref-3}
The hysteresis direction and resistance switching between LRS and 
HRS are the same as the present case; reverse and forward biases 
turn the junction to HRS and LRS, respectively. 
Compared with Ag/PCMO\cite{ref-2} or Ti/TiO$_{x}$/PCMO\cite{ref-3} interfaces, 
the present interface is much better defined. 
The existence of resistance switching action should be related 
with some intrinsic mechanisms. 
We speculate the charging effect at the interface plays a role. 
By the voltage stresses in forward or reverse directions, 
the distribution of trapped charge would be altered, 
resulting in the modification of band lineup or tunneling probability. 
To explain retention or memory effects, one may have to 
take into account a self-trapping effect or configuration interaction, 
by which the trapped charge modifies a microscopic structure of bond 
arrangement to stabilize the trapped state itself.

In summary, we have found that epitaxially and coherently grown 
heterojunction of SRO/Nb:STO shows rectifying $I$-$V$ characteristics 
agreeing with that of a Schottky junction composed of a deep 
work-function metal and an $n$-type semiconductor. 
Well-developed hysteresis is formed so as to alter the junction 
resistance lower (higher) upon forward (reverse) bias stress. 
Presence of the resistance switching effect in heteroepitaxial 
structure should give a due to elucidation the mechanism and also 
to enhancement of the device performance as memories in terms of atomic 
scale interface engineering.\cite{ref-14} 

We would like to thank C.U. Jung, H. Yamada, and J. Matsuno for 
collaboration, and T. Shimizu, I.H. Inoue, and A. Odagawa for 
useful discussions.



\begin{figure}
\caption{Rectifying $I$-$V$ characteristics of a SRO/Nb:STO junction measured 
at 300 K (a solid line) and 50 K (a dashed line). 
The upper panel of insets shows schematic electronic band diagrams 
of a Schottky junction, where $\phi_{\rm B}$ stands for barrier height. 
The left is under forward (negative) bias and the right is 
under reverse (positive) bias. 
The lower inset shows the sample structure and measurement configuration.
\label{fig-1}}
\end{figure}

\begin{figure}
\caption{(Color online) 
Hysteretic $I$-$V$ characteristics of a SRO/Nb:STO junction measured 
with different spans of voltage scan. 
Bias voltage was swept as (a) 
-1.4 V $\to$ 0 V $\to$ $V_{\rm max}$ $\to$ 0 V $\to$ -1.4 V 
with $V_{\rm max}$ varied as 1, 2, 3, and 4 V, 
and (b) 
4 V $\to$ 0 V $\to$ $V_{\rm min}$ $\to$ 0 V $\to$ 4 V 
with $V_{\rm min}$ varied 
as -0.8, -1.0, -1.2, and -1.4 V. 
In both cases, chronological sequence of $I$-$V$ curves is represented 
from (1) to (4).
\label{fig-2}}
\end{figure}

\begin{figure}
\caption{(a) Forward bias $I$-$V$ characteristics of a SRO/Nb:STO junction 
measured at 300 K (solid line) and 50 K (dashed line). 
The filled and open symbols (stars and triangles) represent the high 
and low resistance states ($R_{\rm H}^0$ and $R_{\rm L}^0$), respectively. 
(b) and (c): Resistance switching behaviors at 50 K and 300 K, respectively. 
Pulsed voltage stresses of $V_{\rm p}$ = $\pm$10 V were applied to the junction 
reset to the high resistance state ($R_{\rm H}^0$) at the initial stage 
(before Time = 0), and the change of the resistance was recorded as 
a function of time for different pulse voltage durations 
($\tau_{\rm p}$) of 1 $\mu$s, 100 $\mu$s and 10 ms. 
The resistance values are measured at $V_{\rm bias}$ as shown in (a).
\label{fig-3}}
\end{figure}

\end{document}